\documentclass{aa}  

\usepackage{bbm}
\usepackage{graphicx}
\usepackage[varg]{txfonts}
\usepackage[colorlinks=true, linkcolor=blue, urlcolor=blue, citecolor=blue]{hyperref}
\usepackage[capitalize]{cleveref}

\crefname{section}{Sect.}{Sects.}
\Crefname{section}{Section}{Sections}

\crefname{figure}{Fig.}{Figs.}
\Crefname{figure}{Figure}{Figures}

\crefname{equation}{Eq.}{Eqs.}
\Crefname{equation}{Equation}{Equations}

\begin{document}

\title{\texttt{aim-resolve}: Automatic Identification and Modeling for Bayesian Radio Interferometric Imaging}

\author{
   Richard Fuchs \inst{1} 
   \and 
   Jakob Knollmüller \inst{1,2}
   \and 
   Jakob Roth \inst{3, 4}
   \and 
   Vincent Eberle \inst{3,5}
   \and
   Philipp Frank \inst{3,6}
   \and
   Torsten A. Enßlin \inst{3, 5,7}
   \and 
   Lukas Heinrich \inst{1}
}

\institute{
   Technical University of Munich, TUM School of Natural Sciences, Boltzmannstr. 2, 85748 Garching, Germany \\
   \email{richard.fuchs@tum.de}
   \and
   Department of Astrophysics, Radboud University, Heyendaalseweg 135, Nijmegen, 6525 AJ, Netherlands
   \and
   Max Planck Institute for Astrophysics, Karl-Schwarzschild-Str. 1, 85748 Garching, Germany
   \and
   Technische Universität München (TUM), Boltzmannstr. 3, 85748 Garching, Germany
   \and
   Faculty of Physics, Ludwig-Maximilians-Universität (LMU), Geschwister-Scholl-Platz 1, 80539 Munich, Germany
   \and
   Kavli Institute for Particle Astrophysics \& Cosmology (KIPAC), Stanford University, CA 94305, Stanford, USA
   \and
   German Center for Astrophysics, Postplatz 1, 02826 Görlitz, Germany
}

\date{Received XXXX; accepted XXXX}

\abstract{
Modern radio interferometers deliver large volumes of data containing high-sensitivity sky maps over wide fields-of-view. These large area observations can contain various and superposed structures such as point sources, extended objects, and large-scale diffuse emission. To fully realize the potential of these observations, it is crucial to build appropriate sky emission models which separate and reconstruct the underlying astrophysical components. We introduce \texttt{aim-resolve}, an automatic and iterative method that combines the Bayesian imaging algorithm \texttt{resolve} with deep learning and clustering algorithms in order to jointly solve the reconstruction and source extraction problem. The method identifies and models different astrophysical components in radio observations while providing uncertainty quantification of the results. By using different model descriptions for point sources, extended objects, and diffuse background emission, the method efficiently separates the individual components and improves the overall reconstruction. We demonstrate the effectiveness of this method on synthetic image data containing multiple different sources. We further show the application of \texttt{aim-resolve} to an L-band (856 - 1712 MHz) MeerKAT observation of the radio galaxy ESO 137-006 and other radio galaxies in that environment. We observe a reasonable object identification for both applications, yielding a clean separation of the individual components and precise reconstructions of point sources and extended objects along with detailed uncertainty quantification. In particular, the method enables the creation of catalogs containing source positions and brightnesses and the corresponding uncertainties. The full decoupling of sky emission model and instrument response makes the method applicable to a wide variety of instruments or wavelength bands.
}

\keywords{
   Methods: data analysis --
   Methods: statistical --
   Techniques: image processing -- 
   Instrumentation: interferometers
}

\maketitle

\section{Introduction} \label{sec:introduction}

Modern and upcoming radio interferometers, notably the Australian Square Kilometre Array Pathfinder, MeerKAT, and the Square Kilometre Array (SKA) are leading to major improvements in radio astronomical imaging. These arrays are starting to produce large volumes of data containing high-sensitivity sky maps over wide fields-of-view (FOV). These large area observations often contain various and superposed structures such as point sources, extended objects, and large-scale diffuse emission. To fully realize the potential of these observations, it is crucial to build appropriate sky emission models which reconstruct and separate the underlying astrophysical components. 

This is a challenging task due to several reasons: In general, radio interferometric imaging is an underconstrained problem requiring some form of regularization. As the individual sources vary in both type and size, they cannot be sufficiently described by a single sky emission model. In addition, exact locations of point sources and extended objects are often not known beforehand. Finally, the entire imaging process should be fully automated allowing its application to new and unknown datasets. To solve these problems, we combine Bayesian imaging with object identification and source extraction to an automatic and iterative method named \texttt{aim-resolve}.

Determining the locations of individual objects, often referred to as source finding, is a well-known task in radio astronomy. A summary and comparison of different tools can be found in \citet{hopkins2015}. Recent advances in machine learning have enabled further improvements to source finding. In particular, convolutional neural networks (CNNs) have been successfully used to detect point sources \citep{lukic2019b, vafaei2019, tilley2021} or to classify the morphology of radio galaxies \citep{bowles2021, schmidt2022, taran2023}. 

Most of these source finding algorithms run on images reconstructed using some variant of the \texttt{CLEAN} algorithm, originally proposed by \citet{hogbom1974}, which is the commonly employed imaging algorithm in radio interferometry due to its computational efficiency and simplicity. Although being significantly improved over the last decades, in particular for diffuse emission, wide-field imaging, and spectral imaging \citep{schwab1983, bhatnagar2004, cornwell2008, offringa2017}, \texttt{CLEAN}-based algorithms still have a couple of drawbacks. They do not output any uncertainty information, they often require manual guidance to ensure their stability, and \texttt{CLEAN} restored images often exhibit limited image quality and resolution \citep{arras2021, terris2022}. In addition, \texttt{CLEAN} can produce imaging artifacts, which often need to be excluded manually before or during the source finding process.

Many other imaging algorithms have been developed to improve these limitations. One large class are convex or non-convex optimization algorithms using sparsity based regularizers, originally proposed by \citet{wiaux2009}. Recent examples of such algorithms \citep{birdi2019, dabbech2022, thouvenin2022} have led to improvements in terms of image fidelity and resolution compared to advanced versions of \texttt{CLEAN}. \citet{repetti2019} proposed some form of uncertainty quantification for these methods. 

A second large class of novel imaging algorithms builds on Bayesian inference. By accessing the posterior distribution, Bayesian imaging algorithms address the uncertainty quantification problem from first principle. Many Bayesian imaging algorithms \citep{cai2018, tiede2022} are based on posterior sampling techniques. The Bayesian imaging algorithm \texttt{resolve}\footnote{\url{https://gitlab.mpcdf.mpg.de/ift/resolve}} \citep{junklewitz2016, arras2019} employed in this work builds on variational inference (VI) for computational efficiency. Recent examples are \citet{arras2022} applying \texttt{resolve} to EHT data and \citet{roth2023} joining Bayesian imaging with direction-dependent calibration. \citet{arras2021} shows the enhancements of \texttt{resolve} compared to \texttt{CLEAN}, such as higher image fidelity, super-resolution and uncertainty quantification of the results. However, the improvements come at the cost of higher computational effort. The recently presented \texttt{fast-resolve} algorithm \citep{roth2024} addresses this problem and significantly reduces the computational complexity compared to classical \texttt{resolve}, enabling Bayesian image reconstruction of large data sets like a MeerKAT wide-field observation. 

\texttt{resolve} is build on the numerical information field theory (\texttt{NIFTy}\footnote{\url{https://github.com/NIFTy-PPL/NIFTy}}) python package \citep{selig2013, nifty5, edenhofer2024}, which relies on accurate variational inference (VI) algorithms yielding posterior samples of the reconstructed quantities. \texttt{NIFTy} allows to build complex forward models and to separate different components during the reconstruction of the data. These features have already been successfully applied to X-ray and $\gamma$-ray observations separating point sources or even extended objects from a diffuse background \citep{scheel2023, westerkamp2024, eberle2024}. Very recently, \citet{guardiani2025} introduced a \texttt{NIFTy}-based method to automatically detect and model point sources in astronomical surveys using model latent-space information and showed its application to X-ray data. However, the FOV and model parameters for the extended objects still have been set manually in these examples. This gets infeasible for a large number of extended components as they can occur in radio wide-field observations.

In \texttt{aim-resolve}, we utilize traditional deep learning and clustering algorithms to detect both point sources and extended sources in radio observations. The method first produces a preliminary reconstruction of the radio sky with \texttt{fast-resolve}. Next, it identifies point sources and extended objects within this reconstruction and uses the available information, such as the localization and brightness of the sources, to build a multi-component sky model. By utilizing different model descriptions for different types of sources, the method successfully separates the individual sources from each other. The following joint optimization of the multi-component sky model with the data improves the overall reconstruction while providing uncertainty quantification of the results. The whole method is executed in an automatic and iterative manner.

The remaining paper is organized as follows: \Cref{sec:imaging} discusses the imaging problem in radio interferometry and the corresponding measurement equation. In \cref{sec:method}, we first describe the underlying tools used for modeling, optimization, and object identification, before we combine them to the iterative method of \texttt{aim-resolve}. In \cref{sec:applications}, we validate the method on synthetic radio data and show its application to a MeerKAT radio wide-field observation.

\section{Imaging in radio interferometry} \label{sec:imaging}

For a general image reconstruction task with additive noise the measurement equation reads
\begin{equation} \label{eq:me}
   d = R\,I + n,
\end{equation}
where $d$ is the observed data, $I$ the to be reconstructed image or signal, $R$ is the signal response mapping the image to the data, and $n$ an additive noise realization. The response function depends on the particular instrument.

In radio interferometry we do not measure the sky brightness distribution $I$ directly. Instead, data from a large number of telescopes is combined to measure Fourier components of the sky brightness, often referred to as visibilities. For an ideal, unpolarized interferometer the visibilities $\mathcal{V}$ are given by the radio interferometric measurement equation \citep{thompson2017}
\begin{equation} \label{eq:rime}
   \mathcal{V}(u,v,w) = R\,I = \int \frac{I(l,m)}{n(l,m)} e^{-2\pi i [ul + vm + w(n(l,m) -1)]} dl \, dm,
\end{equation}
where $I(l,m)$ is the sky brightness distribution, ($l$, $m$, $n(l,m) = \sqrt{1 - l^2 - m^2}$) are the sky coordinates, and ($u$, $v$, $w$) the relative positions of antenna pairs. We further assume uncorrelated Gaussian noise.

By combining \cref{eq:me} and \cref{eq:rime}, one easily can generate model visibilities for given sky brightness distributions and noise statistics. However, directly inverting the operation is impossible as the data is corrupted by noise and measured only at sparse locations in Fourier space. For this reason, several different images might be consistent with the same data. Solving this underconstraint problem requires additional regularization, e.g. through the incorporation of prior distributions.

\section{Method} \label{sec:method}

In this section, we introduce the iterative method of \texttt{aim-resolve}. The general approach has been outlined in \citet{fuchs2025}. We first give an overview over the frameworks and tools used within the method. This includes the setup of a flexible generative model allowing for a full description of the radio sky (\cref{sec:modeling}), followed by the utilized optimization algorithms of \texttt{NIFTy} and \texttt{fast-resolve} (\cref{sec:optimization}). Next, we demonstrate how objects can be identified in a reconstruction using deep learning and clustering techniques (\cref{sec:object_id}), and finally combine everything to the \texttt{aim-resolve} method (\cref{sec:it_procedure}).

\subsection{Modeling} \label{sec:modeling}

\texttt{resolve} infers the posterior probability distribution $P(I|d)$ of the sky brightness conditioned on the measured data, instead of using a point estimate. The posterior distribution can be expressed in terms of the evidence $P(d)$,  the likelihood $P(d|I)$ and the prior distribution $P(I)$ via Bayes' Theorem
\begin{equation} \label{eq:bayes}
    P(I|d) = \frac{P(d|I) P(I)}{P(d)}.
\end{equation}
Because of the size and complexity of our imaging problem, we approximate the posterior using the optimization algorithms of \texttt{NIFTy} and \texttt{fast-resolve}, which are discussed in \cref{sec:optimization}. In the next two subsections we setup a versatile and adjustable model for the full sky brightness distribution and formulate the underlying \texttt{NIFTy} prior models. We furthermore describe the likelihood assumed in resolve.

\subsubsection{Sky description} \label{sec:sky}

As stated before, the observed radio sky often contains various different sources with widely varying physical properties. Therefore, a single sky emission model is not sufficient to describe all structures present in radio observations.

Especially diffuse emission and point sources have a very different morphology. For this reason, we use different prior models for point sources and extended objects. Previous applications of \texttt{resolve} followed similar approaches. For example, \citet{arras2021} introduced a separate model for two point sources with known positions to improve the imaging of Cygnus A. In this work, we extend this ansatz to an arbitrary number of point sources whose locations are determined throughout our method.

Besides the separation between extended objects and point sources discussed above, there is also a wide variety of extended sources. To address this, we introduce so-called tile components, pointing to specific sub-parts of the signal field. Additionally, we incorporate a diffuse background component to capture remaining objects and large-scale structures that are not represented by one of the tiles. Together with the point sources, the full forward model for the sky brightness distribution reads
\begin{equation} \label{eq:s_sum}
    I = I^{\text{b}} + \sum_i^N I^{\text{t}}_{i} + \sum_j^M I^{\text{p}}_{j},
\end{equation}
where $ I^{\text{b}}$ represents the diffuse background, $I^{\text{t}}_{i}$ denotes $N$ individual tile components, and $I^{\text{p}}_{j}$ signifies the $M$ point sources. By pairing this model description with sensible prior distributions for the individual components, we can describe various structures and sources at different scales.

\subsubsection{Prior models} \label{sec:prior}

We employ \texttt{NIFTy} to setup the prior models for the three different types of components. A central requirement for all prior models is positivity, as the observed flux in radio astronomy is strictly positive. Additionally, they must be able to model significant variations in sky brightness, which can span several orders of magnitude.

Nearly all continuous quantities, such as galaxies, and large-scale diffuses emission, can be described as correlated structures. To capture this behavior, we describe these structures using Gaussian processes (GPs), which model correlations between nearby pixels represented by a power spectrum in Fourier space. As the exact correlation structure of a component is not known a priori, we use a nonparametric model for this power spectrum and infer it jointly with the GP. Further details of this correlation model can be found in \cite{arras2022}.

In general, we encode our prior models as standardized generative models \citep{knollmuller2018}, which map a set of standard normal distributed random variables $\{\xi\}$ to the desired target distribution. To ensure the non-negativity of the diffuse model, we simply exponentiate the GP model to encode a correlated log-normal distribution
\begin{equation}
    I^{\text{d}}(\xi) = \exp{(\mathcal{F}^{-1} [\sqrt{P_\Psi (\xi_\Psi) }~ \xi_k ] )},
\end{equation}
where $\mathcal{F}^{-1}$ is the inverse Fourier transform operator, and $P_\Psi(\xi_\Psi)$ is the non-parametric power spectrum modeling the spatial correlations. 

Since this model is quite flexible, we can utilize it to describe background and tile components by adjusting the hyper-parameters. Thus, we can express each compact object with its individual correlation structure. For the tile components, we further multiply the output of the log-normal correlation model with a two-dimensional Gaussian with a learnable covariance, similar to the prior model of \citet{ruestig2024}. This reduces the degeneracies in parameter space between the background and the tile components, and ensures that the tile components fall off to zero towards the edges. These tile components are then inserted to the specific sub-parts of the signal field.

While the correlation model works well with different kinds of extended objects, it has problems to describe point sources. In this work, we aim to predict the exact pixel locations of point sources throughout our method. Moreover, we are able to determine the flux of each detected point source. Thus, instead of having a point source prior everywhere in the image space, which would lead to a high degeneracy, we insert an individual model for each identified source $j$ at its predicted pixel coordinates and describe its brightness with a log-normal distribution
\begin{equation}
    I^{\text{p}}_{j}(\xi) = \exp(\mu_{j} + \sigma_{j} \, \xi^{\text{p}}_{j} ),
\end{equation}
which maintains the necessary positivity constraint. It transforms the standard normal distributed random variables $\xi^{\text{p}}_{j}$ to a normal distribution with mean $\mu_{j}$ and standard deviation $\sigma_{j}$.

\subsubsection{\texttt{resolve} likelihood} \label{sec:lh}

Following \citet{arras2021}, we assume Gaussian noise statistics with diagonal covariance $N$, $n \curvearrowleft \mathcal{G}(n,N)$. The likelihood is therefore Gaussian as well, given by
\begin{equation}
    P(d|I) = \mathcal{G}(d - R\,I\,,\, N).
\end{equation}
For numerical reasons, we work with the negative log-likelihood,
\begin{equation} \label{eq:ham}
    H(d|I) = \frac{1}{2} (d - R\,I)^\dagger N^{-1} (d - R\,I) + \text{const},
\end{equation}
with $\dagger$ denoting the complex conjugate transpose. Note that $I$ is the full sky brightness model specified in \cref{eq:s_sum} and may consist of a varying number of different components during the iterative method.

\subsection{Optimization} \label{sec:optimization}

Given the full model for the sky brightness distribution and the likelihood, we want to estimate the posterior distribution in \cref{eq:bayes}. For radio interferometric data with millions of visibilities, we employ the \texttt{fast-resolve} optimization scheme, presented in \cref{sec:fast_resolve}. This algorithm relies on accurate VI algorithms of \texttt{NIFTy}, which are introduced first.

\subsubsection{\texttt{NIFTy} posterior inference} \label{sec:nifty_inf}

In VI, we approximate the posterior distribution using a family of distributions by minimizing the Kullback-Leibler divergence between the approximation and the true distribution. In this work, we employ geometric Variational Inference (geoVI, \citealp{frank2021}) to efficiently approximate the posterior distribution with a set of samples allowing to produce reliable uncertainty estimates for the results. 

In general, VI methods scale much better with the number of parameters compared to sampling techniques like MCMC and HMC, making them still applicable to high resolution imaging problems \citep{blei2017}. However, they still involve a lot of likelihood calls within the inference, which increases the computational costs if the evaluation of the likelihood is expensive. In \texttt{resolve}, each evaluation of the likelihood requires the computation of $RI$ and thus the integral in \cref{eq:rime}. In practice, this boils down to computing a discrete non-uniform Fourier transformation. Although \texttt{resolve} relies on the parallelizable \texttt{wgridder} implemented in the \texttt{ducc}\footnote{\url{https://gitlab.mpcdf.mpg.de/mtr/ducc}} library to perform this operation, the evaluation of $RI$ becomes computationally expensive for large datasets with millions of visibilities.

\subsubsection{\texttt{fast-resolve} inference} \label{sec:fast_resolve}

To eliminate this bottleneck, \citet{roth2024} introduced the \texttt{fast-resolve} algorithm with the aim to reduce the number of response evaluations. The basic idea is to perform most of the computations in image space by multiplying both sides of \cref{eq:me} with $R^{\dagger} N^{-1}$ from the left. By approximating the response function of the new likelihood by a convolution with the point spread function of the interferometer and the application of the noise inverse with an appropriate noise kernel, the evaluation time of the new likelihood can be reduced significantly.

Inspired by the \texttt{CLEAN} algorithm, \texttt{fast-resolve} uses a major/minor cycle inference scheme. In the minor cycles, it optimizes the current estimate of the posterior distribution of the sky brightness using the above approximations. The major cycles then correct for approximation errors by applying the exact response function. For more detail on the exact inference scheme and the gained speedup refer to \citet{roth2024}.

\subsection{Object identification} \label{sec:object_id}

In order to build a full multi-component sky model, we need to predict the locations of both point sources and extended objects in a \texttt{resolve} reconstruction. There are three major challenges in this task: first, the number of objects is unknown before; second, we strive for a highly precise localization of the individual objects, especially of the point sources; and third, the prediction should work on inputs with variable resolutions.

\subsubsection{Semantic segmentation} \label{sec:unet}

We tackle these problems using semantic segmentation, which provides a detailed understanding of the image’s content at the pixel level by assigning labels to every pixel. We chose the U-Net architecture \citep{ronneberger2015} for our prediction network, which features an encoder-decoder structure with skip connections enabling precise localization of features while maintaining spatial details. Due to its fully-convolutional architecture, it can even process inputs with different resolutions. 

We set up the network in a way that it takes a reconstructed image as input and produces two output segmentation maps, indicating whether a pixel represents a point source or not, or if it belongs to an extended object or not. After training the U-Net on synthetic image and label pairs, we can get fast predictions of the source locations when running our iterative method.

\subsubsection{Training data} \label{sec:train_data}

For the training data, we set up a generative model to produce synthetic images and labels at various resolutions. The images contain plenty of different structures to resemble actual \texttt{resolve} reconstructions of radio wide-field observations. 

To achieve this, we introduce some diffuse background emission in the middle of the FOV by multiplying the lognormal correlation model from \cref{sec:prior} with a two-dimensional Gaussian. We generate extended objects that look like real radio galaxies by using part of the \texttt{RadioGalaxyDataset}\footnote{\url{https://zenodo.org/records/7692494}}, a collection of several radio catalogues \citep{becker1995, miraghaei2017, gendre2008, gendre2010, capetti2017i, capatti2017ii, baldi2018, proctor2011}, and add them to the background. We further add variable numbers of point sources and small extended objects at random locations. The point sources can be blurred over several pixels, since our correlation model cannot perfectly represent them. At the same time, we generate the two output segmentation maps using the pixel locations of the inserted point sources and compact objects, as well as the pixels occupied by the extended objects from the \texttt{RadioGalaxyDataset} images. A subset of the training data can be found in \cref{fig:train_data}.

\subsubsection{Clustering} \label{sec:clustering}

From the point source segmentation map, we can directly extract the exact pixel locations of the sources. However, we only get a single output map which can contain several extended objects. To discriminate between them, we employ DBSCAN (Density-Based Spatial Clustering of Applications with Noise) introduced by \cite{ester1996}. Compared to other clustering algorithms it does not require the number of clusters as input and works well for data with clusters of similar density, which is the case for two-dimensional images. By setting a minimum cluster size, DBSCAN can additionally have a regulating role, e.g. it can avoid that blurred-out point sources are miss-classified as small extended objects.

\begin{figure*}
\centering
   \includegraphics[width=17cm]{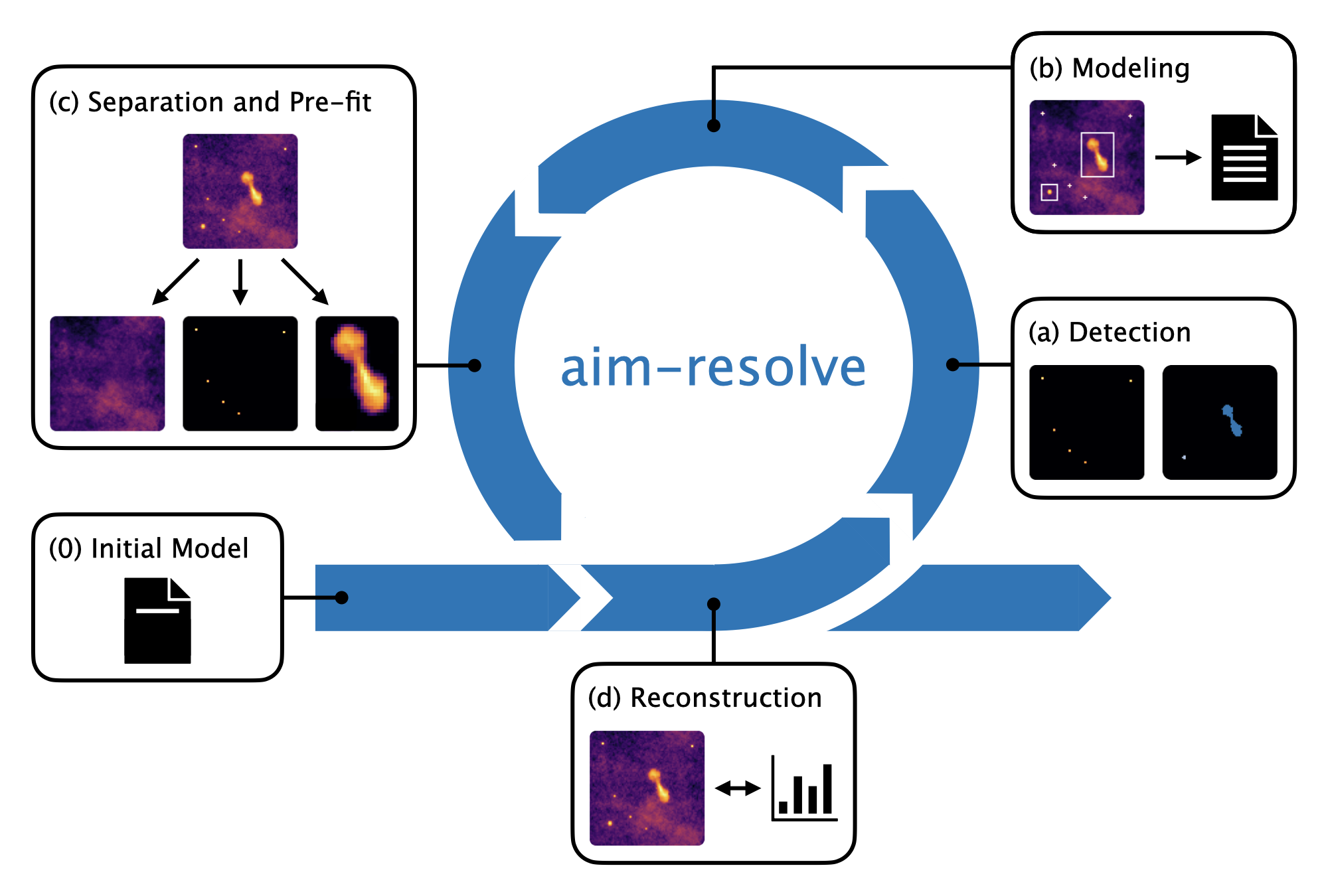}
     \caption{Visualization of the iterative method of \texttt{aim-resolve}. Initialized with a single component model (0), it generates a preliminary reconstruction of the data (d). It detects point sources and extended objects in the reconstruction (a) and adds the identified components, marked with crosses and boxes, to the existing background model (b). Next, it separates the components from the background and fits them to the previous reconstructed image (c), before it continues the optimization on the real data (d). This cycle is repeated multiple times for further refinement.}
     \label{fig:procedure}
\end{figure*}

\subsection{Iterative procedure} \label{sec:it_procedure}

We combine the modeling, optimization, and identification steps to the method \texttt{aim-resolve} in order to iteratively improve the reconstruction of the data. We set up a fully automatized pipeline using the \texttt{snakemake} workflow language \citep{molder2021} to efficiently manage the sequence of the individual steps.

\Cref{fig:procedure} shows a visualization of \texttt{aim-resolve}, depicting the individual steps of the method. In step (0) we initialize the process, load the data and specify the desired resolution and FOV we wish to reconstruct. On the model site, we start with a single component model, which consists solely of the background component from \cref{sec:sky}, and initialize its correlation model with rather wide and flexible hyper-parameter priors. After running the inference in step (d), we get a preliminary reconstruction of the data that serves as a starting point for our method, which iterates over the following steps:

\begin{itemize}
    \item[(a)] \textit{Detection}: 
    First, the trained U-Net is used to identify point sources and objects within the preliminary reconstruction. The point source positions can be extracted directly from the corresponding output segmentation map. To discriminate between the different extended objects, the clustering algorithm from \cref{sec:clustering} is applied to the corresponding output map. \vspace{6pt}
    
    \item[(b)] \textit{Modeling}: 
    This step determines the center of the contained object for each output cluster and draws a bounding box by adding some margin around it. Additionally, it extracts the flux of each point source and object from the previous reconstruction and sets the new model hyper-parameters accordingly. Using all the available information, it creates a model configuration file for the subsequent reconstruction iteration by adding the new components to the background model, yielding a multi-component model. \vspace{6pt}
    
    \item[(c)] \textit{Separation and Pre-fit}:
    Before continuing the optimization on the data, the new model is fitted to the previous reconstructed image. This step is similar to the transition model introduced by \citet{westerkamp2024}. By masking the background at the locations of the detected components, this step efficiently separates the point sources and extended objects from the background. Next, the individual model components are fitted to the corresponding areas of the old reconstruction. Note that the transition saves computation time since the pre-fit on the previous reconstructed image does not involve any expensive response function. More details on this transition to the new model can be found in \cref{app:transition}. \vspace{6pt}
    
    \item[(d)] \textit{Reconstruction}:
    The pre-fitted model serves as a starting point to continue the optimization on the data. With the new multi-component model, we now can create plots for each individual component separated from each other. Moreover, the components can be added together to compose an image of the full sky, allowing for object detection in the next pipeline iteration.

\end{itemize}
The iterative pipeline is repeated multiple times for further refinement and stops once the reconstruction does not improve anymore. Note that we re-predict all point sources and extended objects at each detection iteration. This ensures that potentially wrongly identified components from the previous detection step are corrected or deleted and thus the overall reconstruction iteratively improves.

\section{Applications} \label{sec:applications}

We now demonstrate the applicability of our method. We first describe the training of the U-Net and its performance on unseen test data in \cref{sec:training}. In \cref{sec:validation}, we validate our method on synthetic radio data before we finally apply it to real radio interferometric data in \cref{sec:meerkat}.

\subsection{Training of the U-Net}\label{sec:training}

We use the generative model introduced in \cref{sec:train_data} to produce several train, validation, and test datasets at various resolutions. All of them can contain some diffuse background emission, large radio-galaxy-like objects and up to 25 point sources and small extended components. The idea of differently resolved datasets is to first train the U-Net described in \cref{sec:unet} on a dataset with a rather low resolution, which reduces the memory usage and allows faster training, and later fine-tune the U-Net on train data with a high resolution to achieve a good performance on high resolution images. 

Before the data is fed into the U-Net, we take the logarithm of the images and normalize them. Furthermore, we randomly flip and rotate the images and labels to allow different orientations of the radio galaxy-like objects within the images. We set up the U-Net using the implementation from \citet{Iakubovski2019} with a ResNet-50 encoder pre-trained on the ImageNet\footnote{\url{https://image-net.org}} data set, with one input channel and two output channels. We achieved best results using a dice loss function, Adam optimizer, and Cosine Annealing learning rate scheduler. For the first step, we train the U-Net on a dataset containing 10000 images and labels with a resolution of $128 \times 128$. Afterwards, we fine-tune the trained U-Net on the final resolution the neural network should operate on. We do this for two different resolutions, $512 \times 512$ and $1024 \times 1024$, with 1000 images and labels each.

To measure the performance of the network's predictions, we use the mean intersection of union (mIoU). Unlike the pixel-wise accuracy, mIoU aims to mitigate class imbalance and to take false positives into account. However, the per-dataset mIoU is biased towards large objects in the dataset. Our dataset shows exactly this size imbalance, since extended objects contain more pixels than point sources. \citet{wang2023} propose using fine-grained mIoU at per-image level to address this problem. We first compute the mIoU score for each image $i$ and each class $c$ via
\begin{equation} \label{eq:miou}
    \mathrm{IoU}_{i,c} = \frac{\mathrm{TP}_{i,c}}{\mathrm{TP}_{i,c} + \mathrm{FP}_{i,c} + \mathrm{FN}_{i,c}},
\end{equation}
where TP, FP and FN represent true positives, false positives and false negatives, respectively. Next, we average these scores first by image and then by class. The order of the averaging ensures that point sources and extended objects are given the same importance. \Cref{tab:unet} compares the fine-grained mIoU at per-image level of several fine-tuned U-Nets on test datasets with different resolutions.

\begin{table}[h]
\caption{Performance of the U-Net}
\label{tab:unet}
\centering
\begin{tabular}{c c c c}
\hline\hline
Dataset & U-Net 128 & U-Net 512 & U-Net 1024 \\
\hline 
    test-128 & 0.937 & 0.891 & 0.716 \\
    test-512 &  0.887 & 0.938 & 0.869 \\
    test-1024 & 0.539 & 0.721 & 0.927 \\
\hline
\end{tabular}
\end{table}

As expected, the U-Nets perform best on the resolution they are fine-tuned on. U-Net 512 and U-Net 1024 still perform acceptable on lower-resolution inputs due to the pre-training. Although the fine-tuning requires additional training steps, it considerably improves the performance with a fairly short additional training time. Of course, the training time increases when going to even higher resolutions. 

Compared to just identifying the brightest pixels in a reconstruction by setting some cutoff value, the U-Nets can separate point sources from extended objects, which is not trivial for this dataset since a varying diffuse background can have a similar brightness as some of the objects and point sources can be blurred out.

\subsection{Validation on synthetic data} \label{sec:validation}

To evaluate the performance and robustness of our proposed method, we validate the reconstruction pipeline on synthetic radio data. The main objectives of this validation are: to verify that the combination of U-Net and clustering efficiently identifies and distinguishes different objects within a reconstruction. And that the multi component model (MCM) works in combination with a radio interferometric response and improves the overall reconstruction quality compared to a single component model (SCM).

\begin{figure*}
\centering
   \includegraphics[width=17cm]{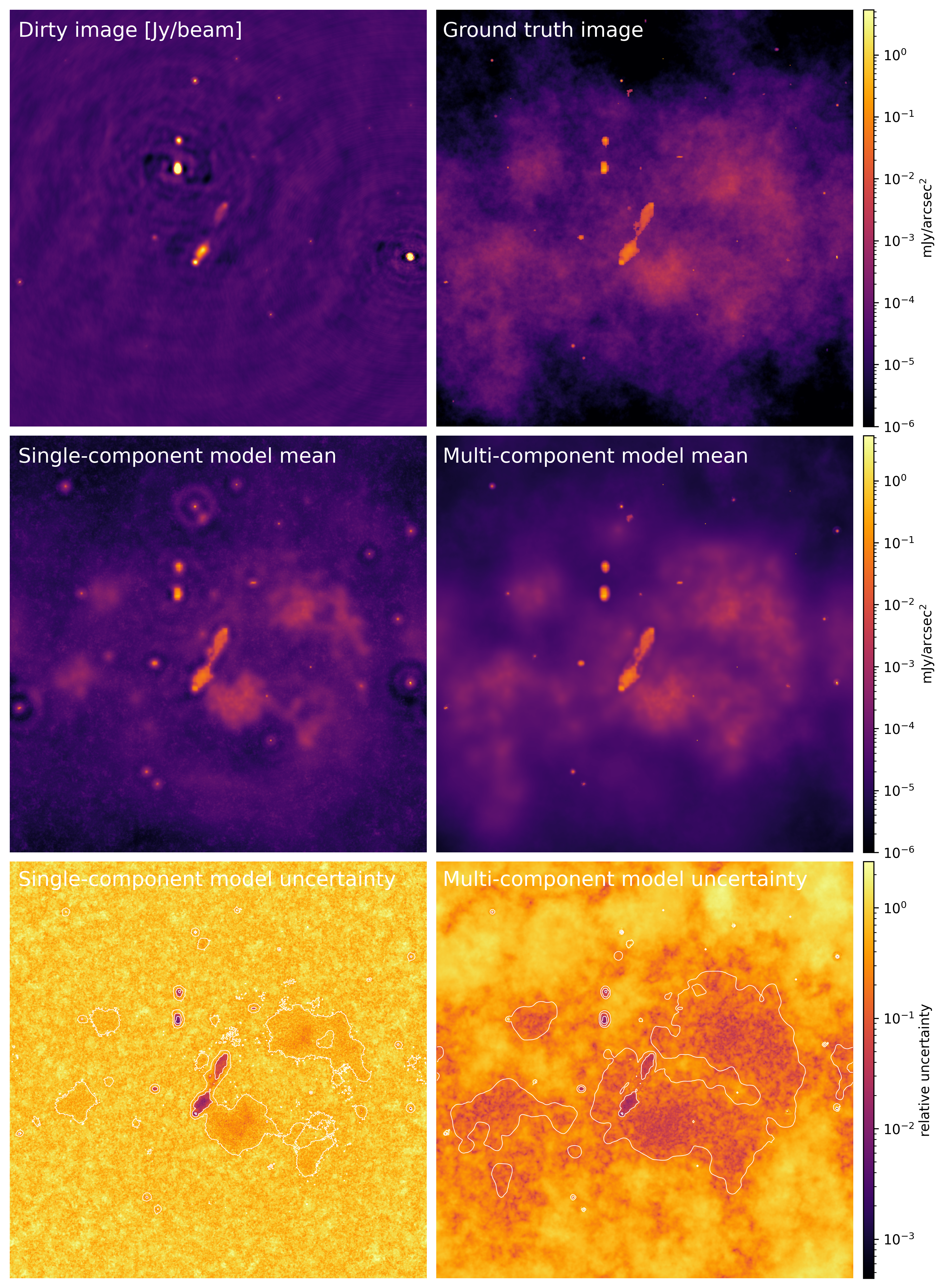}
     \caption{Comparison of the reconstructions of the synthetic radio data using the single-component model (SCM) and the multi-component model (MCM, after the 3rd iteration of \texttt{aim-resolve}). The first row shows the naturally weighted dirty image of the radio data and the underlying ground truth image. The following rows illustrate posterior mean and relative standard deviation of the reconstructions with the SCM and MCM after full convergence, respectively. The MCM reconstruction shows more precise and detailed reconstructions for the point sources and small extended objects, which is further illustrated by tighter flux contour lines around the sources in the MCM uncertainty map.}
     \label{fig:exp_recs}
\end{figure*}

For the synthetic data, we pick one sample from the test-512 dataset, apply the MeerKAT response function from \cref{sec:meerkat} to it, and add Gaussian noise to the generated visibilities. The ground truth image is shown in \cref{fig:exp_recs}. The selected example contains some diffuse background emission and two large radio-galaxy-like objects with two lobes each, as well as 12 point sources and 25 small extended objects. \Cref{fig:exp_recs} further shows the naturally weighted dirty image, which we get by applying $R^{\dagger} N^{-1}$ to the generated radio data. This serves as a starting point for the \texttt{aim-resolve} inference.

\subsubsection{Reconstruction and identification}

As described in \cref{sec:it_procedure}, we first get a preliminary reconstruction of the noisy image data using the SCM and the \texttt{fast-resolve} algorithm. The initial hyper-parameters of the resolve correlation model are given in \cref{tab:params}. Using the U-Net fine-tuned on the test-512 data and following the detection and modeling steps of \cref{sec:it_procedure}, \texttt{aim-resolve} identifies and distinguishes several components within the reconstruction and draws bounding boxes around the predicted extended objects. Note that we set a fixed minimal component shape of $32 \times 32$ pixel for the small extended objects to increase the efficiency of the model. Consequently, most of the small extended components end up having the same shape and field of view.

\begin{figure*}
\sidecaption
  \includegraphics[width=12cm]{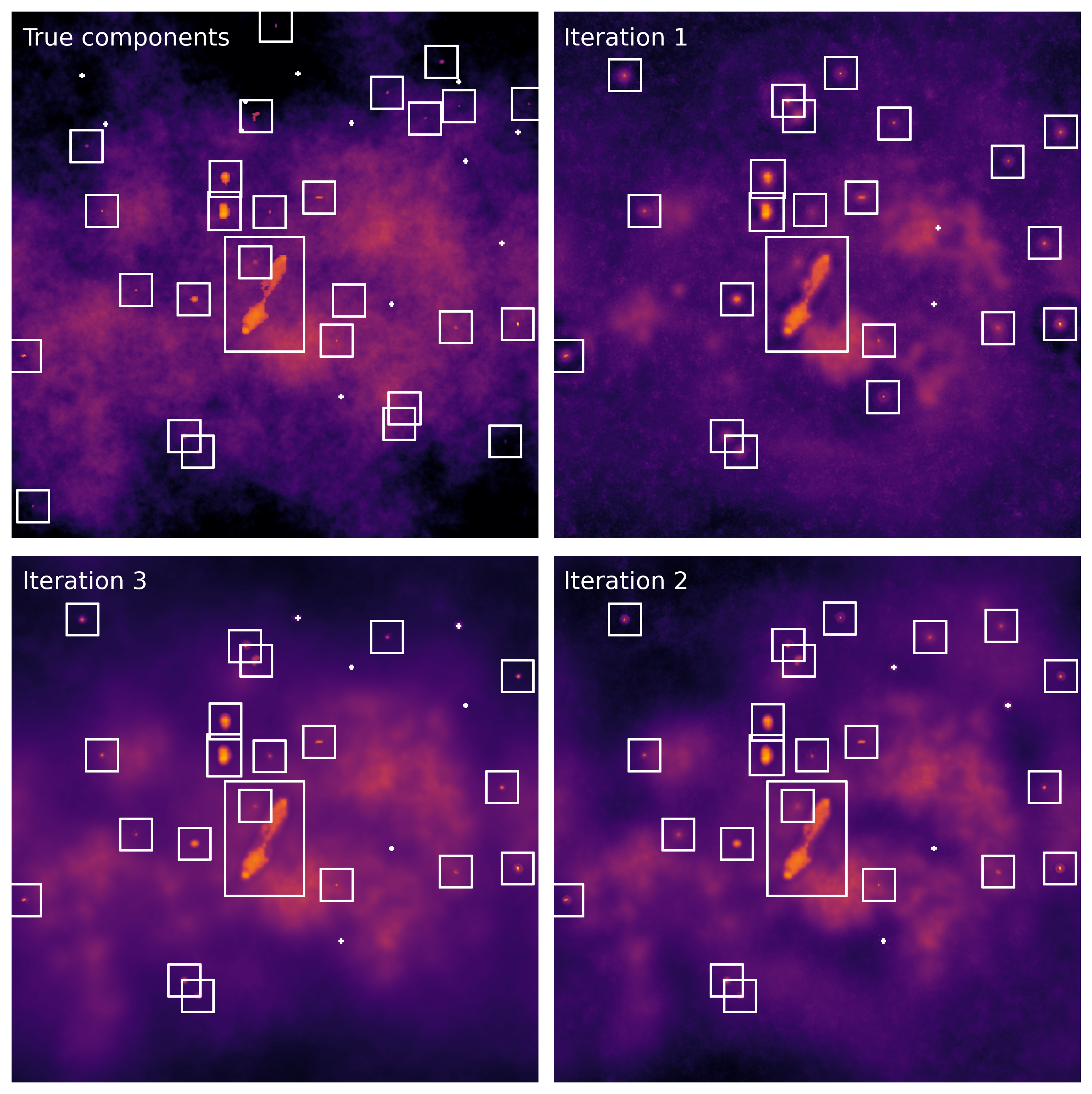}
     \caption{Identified components for several iterations of \texttt{aim-resolve} applied to synthetic image data. The left top image shows the ground truth along with the true components. The remaining images illustrate the reconstructed posterior means for iterations 1 to 3, together with detected point sources (crosses) and extended objects (boxes), showing an increasing number of identified components and an enhancement in overall reconstruction quality across the iterations.}
     \label{fig:exp_boxes}
\end{figure*}

\Cref{fig:exp_boxes} shows the reconstructed posterior mean plus identified point sources (crosses) and extended objects (boxes) for several iterations of the method. These are compared with the ground truth data and true components, which we get by applying the clustering and modeling steps of our method to the true segmentation maps of the generated data. The SCM already manages to reconstruct nearly a dozen of objects. This reconstruction serves as the starting point for several iterations of \texttt{aim-resolve}, which in the end identifies all sources visible in the reconstruction.

Moreover, the detection improves with the quality of the reconstruction, allowing to better distinguish between point sources and extended objects and getting closer to the true components. However, it does not manage to detect all of the sources, especially if the flux of the sources is not much higher than the background, and it miss-classifies some of the point sources as small extended objects. In general, modeling a point source as a small object is not a big problem, since the introduced component can have a different underlying spatial power spectrum than the background, which should improve the reconstruction in the following pipeline iteration. In fact, the re-prediction of the components at each detection step allows that several miss-classified point sources in the first iteration can be identified correctly in a later iteration.

\subsubsection{Model comparison} \label{sec:exp_comp}

When building the MCM, one still has to take care on how to set the priors for the hyper-parameters. For the point sources, we take the total flux from the previous reconstructed image with a wide standard deviation around it. For the correlation models, it is possible to set most of the hyper-parameters quite wide and flexible and just specify the overall offset depending on the previous reconstruction. An overview of the hyper-parameter priors for the different diffuse components is listed in \cref{tab:params}. Recall that the MCM is pre-fitted on the previous reconstruction and further optimized on the real data. This usually results in a full convergence of the model within a couple of major cycles, meaning that the reconstruction does not improve anymore.

\Cref{fig:exp_recs} shows the reconstructed posterior mean and relative uncertainties of both the MCM (after the 3rd iteration of \texttt{aim-resolve}) and SCM along with the ground truth data. For a fair comparison, we further optimize the SCM after the first iteration of the method until full convergence. The MCM shows a better reconstruction of the diffuse background than the SCM. When looking at the point sources and small extended objects, the SCM shows some ringing effects around the sources. These non-physical effects originate from the excitation of high frequency modes in the power spectrum of the GP model. This enables the model to fit multiple compact sources, but introduces some ringing effects in the image. The MCM does not show this behavior as it describes compact objects by separate correlation models and point sources by a single pixel, which can be recognized by careful inspection of the MCM plot. Additionally, it captures more of the true components present in the data. A comparison of the relative uncertainties confirms that utilizing the MCM improves the description of the individual components and the diffuse background, but also the overall reconstruction of the data.

\subsubsection{Component separation} \label{sec:exp_sep}

\begin{figure*}
\centering
   \includegraphics[width=17cm]{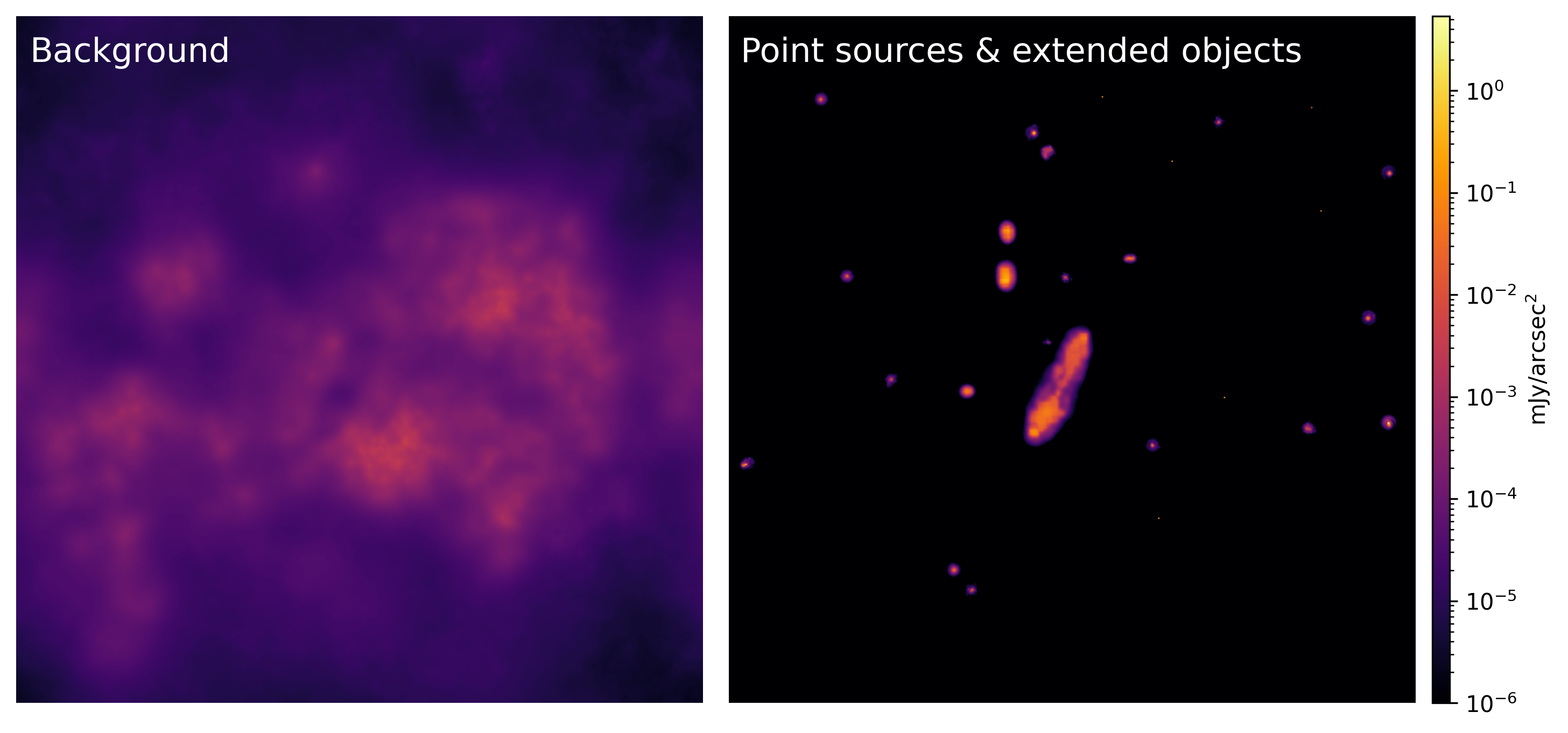}
     \caption{Separate reconstructions of the components of the MCM in \cref{fig:exp_recs}. The left image shows the reconstructed posterior mean of the diffuse background. The right image shows a combined plot of all point sources and extended objects. The sources are well-separated from the background, which captures only the large-scale diffuse emission present in the data.}
     \label{fig:exp_comps}
\end{figure*}

Besides the better reconstruction quality, the MCM efficiently separates point sources and extended objects from the background. \Cref{fig:exp_comps} illustrates the reconstructed posterior means of both the diffuse background and a combined plot of the point sources and extended objects of the MCM. All sources are well-separated from the background, which only shows the diffuse emission present in the data. Note that the MCM further enables to get distinct plots for all individual model components.

\subsection{Application to MeerKAT data} \label{sec:meerkat}

After validating the method on simulated data in the previous section, this section shows the application to a real MeerKAT observation. We apply \texttt{aim-resolve} to an L-band (856 - 1712 MHz) MeerKAT \citep{jonas2016} observation of the radio galaxy ESO 137-006. This dataset offers a great test case as it contains a combination of diffuse emission, point sources, and extended objects, notably the two radio galaxies  ESO 137-006 and ESO 137-007. 

The observation utilized all 64 MeerKAT antennas and the 4k mode of the SKARAB correlator in full polarization with 4096 channels for a total on target time of 14 hours. The novelty of this observation is the first discovery of collimated synchrotron threads connecting the two lobes of a radio galaxy, first presented by \citet{ramatsoku2020}. Since then, the data has been used by \citet{dabbech2022} for demonstrating a sparsity-based imaging algorithm and \citet{roth2024} for presenting the \texttt{fast-resolve} algorithm.

As in \citet{ramatsoku2020}, the data is averaged to 1024 frequency channels and divided into two sub-bands which are relatively free from radio frequency interference, the LO band (961 - 1145 MHz) and the HI band (1295 - 1503 MHz). These two sub-bands are phase self-calibrated using WSClean multi-scale \texttt{CLEAN} and CubiCal for imaging and calibration, respectively. In this work, we image only a subset of 7 frequencies (986 - 1137 MHz) of the LO band with \texttt{aim-resolve}.

\subsubsection{Setup and Identification}

Similarly to the synthetic case, we initialize the SCM with the wide hyper-parameter priors listed in \cref{tab:params} but with a resolution of $1024 \times 1024$ and a FOV of $2\degr \times 2\degr$. Consequently, we choose the U-Net fine-tuned on the \text{test-1024} data with the same resolution for the detection steps. As before we set a fixed minimal shape of $32 \times 32$ for the small extended object components to save computation time.

\begin{figure*}
\sidecaption
  \includegraphics[width=12cm]{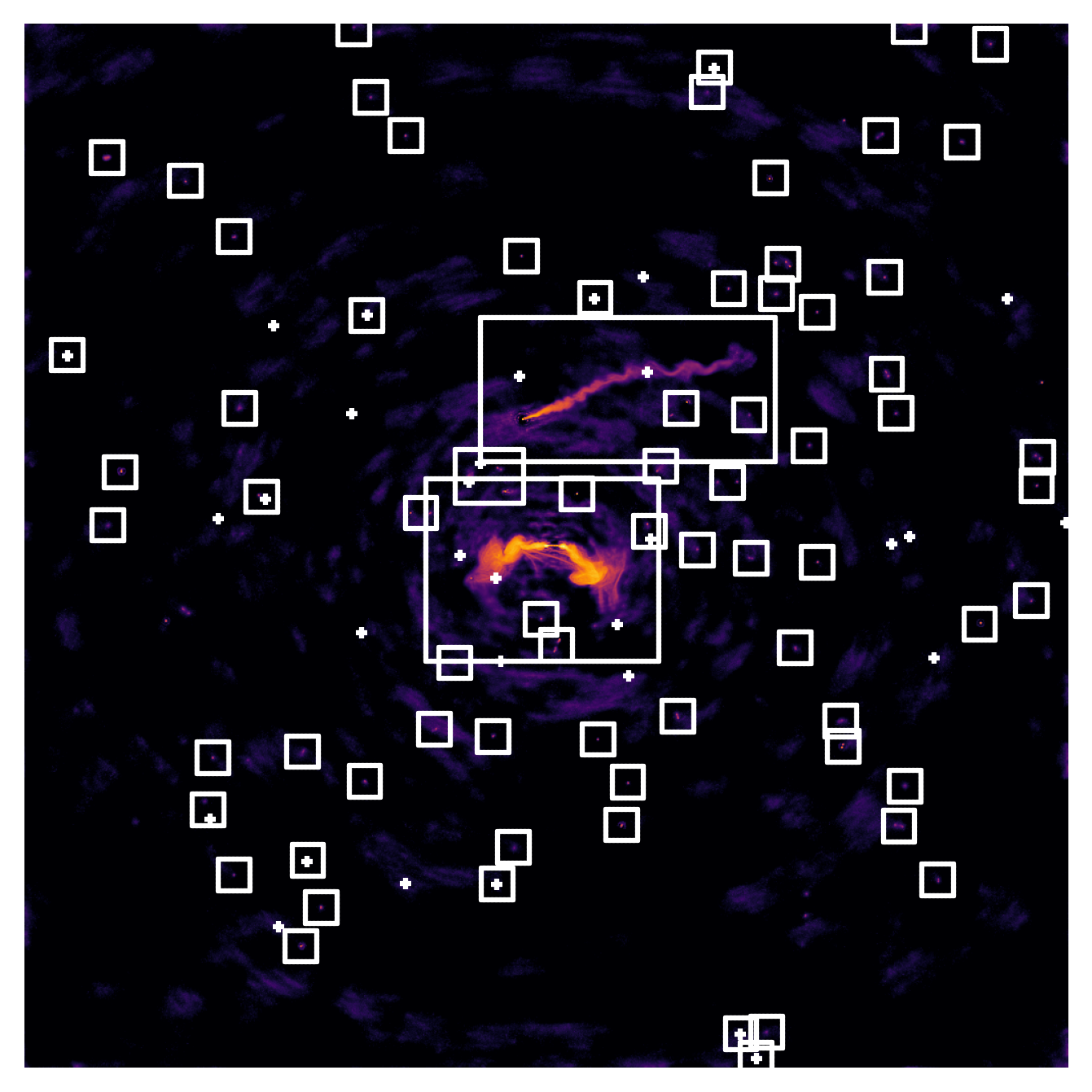}
     \caption{Identified components in the 2nd iteration of \texttt{aim-resolve} applied to the ESO 137-006 MeerKAT observation in the LO sub-band (961 - 1145 MHz) with a FOV of $2\degr \times 2\degr$. The two large boxes indicate the identified galaxies ESO 137-006 and ESO 137-007. The remaining detected objects and point sources are marked with small boxes and crosses, respectively.}
     \label{fig:eso_boxes}
\end{figure*}

We run \texttt{aim-resolve} for several iterations to continuously improve the overall reconstruction and component detection. \Cref{fig:eso_boxes} shows the identified components in the 2nd iteration of \texttt{aim-resolve}. The method detects the ESO 137-006 galaxy in the center of the FOV and the ESO 137-007 wide-tail galaxy on top of it, as well as most of the point sources and small extended objects. In summary, the U-Net detection still works for the MeerKAT wide-field radio observation allowing for an efficient separation of the components in the following.

\subsubsection{Reconstruction} \label{sec:eso_rec}

\begin{figure*}[t]
\centering
   \includegraphics[width=17cm]{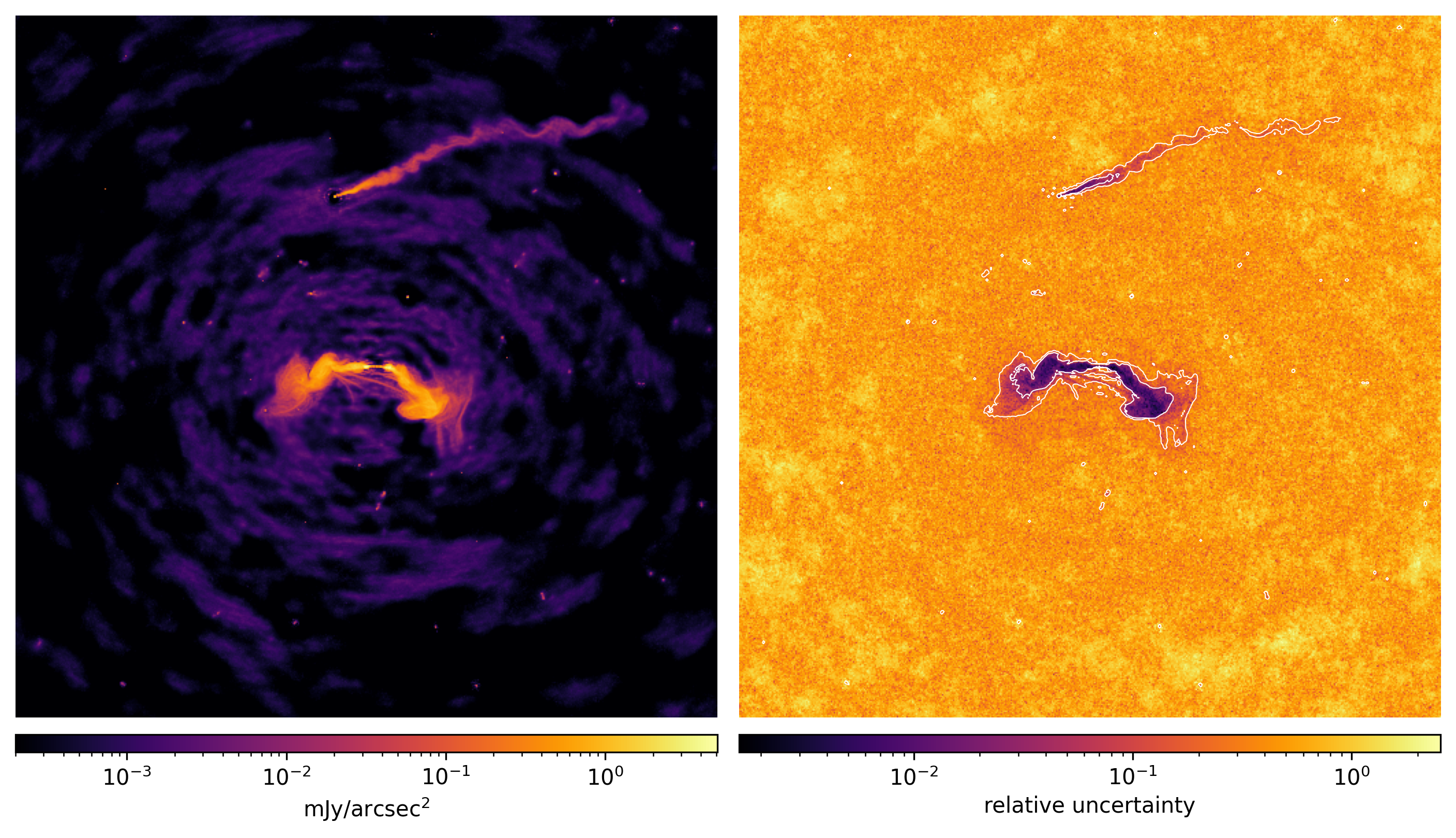}
     \caption{Reconstruction of the ESO 137-006 MeerKAT observation in the LO sub-band (961–1145 MHz) using the MCM (after the 2nd iteration of \texttt{aim-resolve}). The plots show $1\degr \times 1\degr$ zoom-ins of the reconstructed posterior mean and relative standard deviation, covering the two foreground galaxies and many smaller sources. The relative uncertainty is lowest in high-brightness regions (indicated by the flux contours).}
     \label{fig:eso_recs}
\end{figure*}

As before, we set the priors for the MCM's hyper-parameters similarly to \cref{sec:exp_comp}, which can be found in \cref{tab:params}. Note that we need the wide background FOV to capture all relevant sources present in the data. For plotting, however, we focus on the central $1\degr \times 1\degr$ of the observation that contains most of the flux. \Cref{fig:eso_recs} illustrates the resulting reconstructed posterior mean and relative standard deviation within this smaller FOV of the MCM fitted to the MeerKAT data (after the 2nd iteration of \texttt{aim-resolve}). The reconstruction shows precise descriptions of the two foreground galaxies and of many small background sources. Moreover, the MCM captures fine structures like the collimated synchrotron links of ESO 137-006, paired with a high certainty on these sharp structures.

The reconstruction features quite high background artifacts. They are also visible in the \texttt{fast-resolve} reconstructions from \citet{roth2024}. Since the sky prior models in \texttt{resolve} are flexible and therefore sensitive to miscalibrations in the data, this leads to imaging artifacts in the case of suboptimal calibration. Thanks to the multi-component approach, these artifacts get mainly absorbed into the background component and therefore do not affect the reconstructions of the point sources and tile components.

\subsubsection{Component separation} \label{sec:eso_comp}

\begin{figure*}
\centering
   \includegraphics[width=17cm]{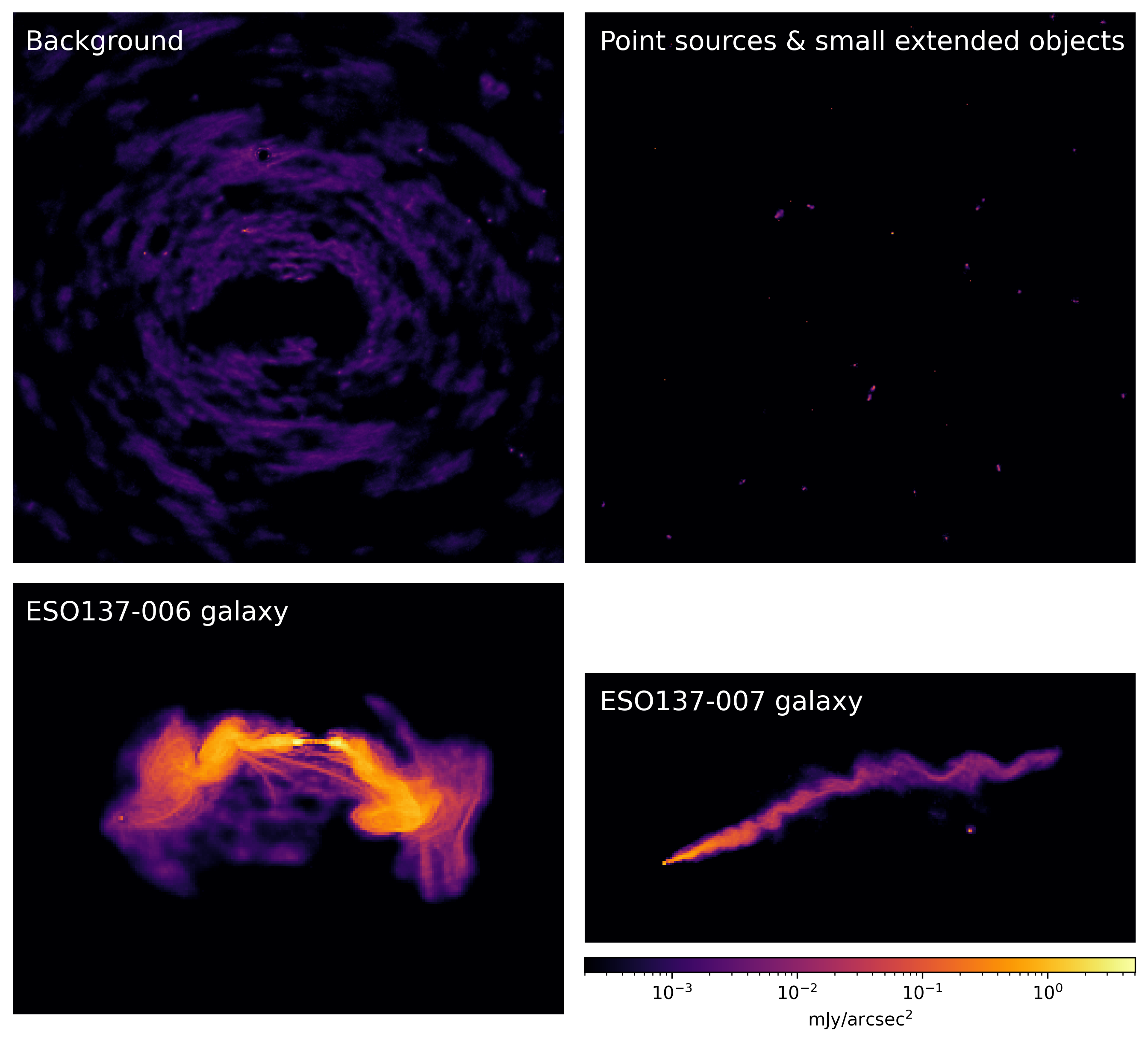}
     \caption{Reconstructions of the individual components of the MCM in \cref{fig:eso_recs}. The images show the reconstructed posterior mean of the background, of the sum of point sources and small extended components, and of the two galaxies ESO 137-006 and ESO 137-007. Most sources are well-separated from the background, which captures only a few other objects besides the diffuse emission and calibration artifacts.}
     \label{fig:eso_comps}
\end{figure*}

Besides a good reconstruction quality, the MCM efficiently separates point sources and extended objects from the background. The first row of \cref{fig:eso_comps} illustrates the reconstructed posterior mean of both the background field and a combined plot of the point sources and small extended objects of the MCM. The second row depicts the two main galaxies ESO 137-006 and ESO 137-007. Because of the many sources present in the data and the suboptimal calibration, the method does not manage to separate all small sources from the background. Nevertheless, the background model has to describe considerably fewer sources than in case of a SCM. 

Lastly, we emphasize that the application of \texttt{aim-resolve} to a large MeerKAT observation implies a trade-off between fidelity and runtime. Higher resolutions of the model would require either more GPU memory or the combination of several GPUs. The primary goal of this paper, however, is to showcase the applicability of the method to real data and not to reach highest reconstruction fidelity.

\section{Conclusions}

In this work, we presented the method \texttt{aim-resolve}, designed to improve imaging for wide-field radio observations containing various different sources. We introduced the individual steps of the method - detection, modeling, separation and pre-fit, and reconstruction - and combined them to an automatized and iterative method. 

To validate the effectiveness of our method, we set up and solved for a generic imaging problem. Within a few iterations, \texttt{aim-resolve} successfully identified all visible point sources and extended objects and efficiently separated them from the background. The utilization of different model descriptions allowed for a more precise reconstruction of the individual components and improved the overall reconstruction of the data. Detailed plots of the reconstructed sky maps demonstrated the advantages of the multi-component approach.

In addition, \texttt{aim-resolve} was applied to an L-band (856 - 1712 MHz) MeerKAT observation of the radio galaxy ESO 137-006 and other radio galaxies in that environment. The method managed to identify and separate the two main galaxies ESO 137-006 and ESO 137-007 as well as many of the small extended objects and point sources. The multi-component reconstruction showed many details, especially in the fine structures of ESO 137-006, and a good description of small sources due to a clean separation of the modeled components.

The separation of the individual components could be used for detailed studies of interesting sources in that environment. Moreover, the gained knowledge on the identified components could be summarized to catalogues containing the positions, brightnesses, and other properties of point sources and extended objects in the FOV, together with the corresponding uncertainties. However, this is beyond the scope of this work.

A possible improvement of the method could be to allow for a higher resolution of the components compared to the background. While this is straightforward to do within \texttt{resolve}, the current implementation of \texttt{fast-resolve} does not support this. Another feature already implemented in \texttt{resolve} is unified imaging and calibration \citep{arras2019}. Adding this feature to \texttt{aim-resolve} could help to remove a large fraction of the imaging artifacts present in the MeerKAT reconstruction. Moreover, it could be beneficial to combine the latent-space based point source detection of \citet{guardiani2025} with the deep learning approach of \texttt{aim-resolve}. 

Interesting features of the ESO137-006 galaxy like the collimated synchrotron links further motivate to extend the multi-component sky model to handle multi-frequency data and to model the spectral behavior of the different astrophysical components. This already has been done for other frequency bands and instruments using \texttt{NIFTy} \citep{scheel2023, westerkamp2024, eberle2024, ruestig2024} and can give interesting insights into the evolution of radio-emitting plasmas in the imaged galaxies.

Finally, the full decoupling of sky emission model and instrument response enables the application of \texttt{aim-resolve} to a wide variety of instruments and wavelength bands, just by an exchange of the response function paired with a slight adaption of the training dataset. This offers the stage for future multi-frequency and multi-instrument reconstructions.

\section{Data availability}

The raw data for the ESO137-006 observation is publicly available via the SARAO archive\footnote{\url{https://archive.sarao.ac.za}} (project ID SCI-20190418-SM-01).
The implementation of \texttt{aim-resolve} will be made publicly available on GitHub\footnote{\url{https://github.com/rifu4/aim-resolve}}.

\begin{acknowledgements}

R.F., V.E., and P.F. acknowledge funding through the German Federal Ministry of Education and Research for the project ErUM-IFT: Informationsfeldtheorie für Experimente an Großforschungsanlagen (Förderkennzeichen: 05D23EO1).

L.H. and J.K. are supported by the Excellence Cluster ORIGINS, which is funded by the Deutsche Forschungsgemeinschaft (DFG, German Research Foundation) under Germany’s Excellence Strategy - EXC-2094-390783311.

J.K. also acknowledges funding from the European Research Council (ERC) under the European Union's Horizon 2020 research and innovation programme (ERC Synergy Grant "BlackHolistic", Agreement No. 101071643).

J.R. acknowledges financial support from the German Federal Ministry of Education and Research (BMBF) under grant 05A23WO1 (Verbundprojekt D-MeerKAT III).

Funded by the European Union. Views and opinions expressed are however those of the author(s) only and do not necessarily reflect those of the European Union or the European Research Council Executive Agency. Neither the European Union nor the granting authority can be held responsible for them. This work is supported by ERC grant (mw-atlas, 101166905).

\end{acknowledgements}

\bibliographystyle{aa}
\bibliography{references}

\appendix

\section{Transitions from previous to new sky model} \label{app:transition}

The main idea of the transition model is outlined in \citet{westerkamp2024} performing a switch from a single-frequency to a multi-frequency model. For our specific problem, we define a transition function that plausible maps the parameters of the sky model of the previous \texttt{aim-resolve} iteration to the model parameters of the new iteration,
\begin{equation}
    T: \xi_\mathrm{prev} \rightarrow \xi_\mathrm{new}.
\end{equation}
Depending on the current iteration of \texttt{aim-resolve}, the previous model can be either a SCM or MCM whereas the new model usually has multiple components. 

We implement the transition by stringing together multiple \texttt{NIFTy} inference problems. First, we create a mask around the predicted point sources and tile components of the new model and fit the masked background component of the new model to the previous reconstruction. With this step, we extract the predicted components from the background. For each predicted point source and tile, we fit a separate model component to the difference of the previous reconstruction and the newly optimized background. Finally, we put together the parameter vectors of the individual components and use this as a starting point for the next reconstruction step on the real data.

Note that we omit any response function during the transition and perform the whole \texttt{NIFTy} optimization in image space. This corresponds to an identity response matrix in \cref{eq:me} and saves computation time.

\section{Prior parameters for the Gaussian process models}

In \cref{tab:params} we list the hyper-parameters (mean and standard deviation) of the Gaussian process model described in \cref{sec:prior}. They are similar for the SCM and the different components of the MCM, only the additional offset parameters of the models differ. For the SCM the offset is set to 12. For the MCM tile components it is set depending on the mean flux in the vicinity of the objects in the previous reconstruction. For the MCM background it is set to the mean flux of the whole reconstruction except the areas around detected components. Note that the remaining hyper-parameters are set quite wide to allow the model parameters to deviate sufficiently from the set prior values. The exact definitions of all model parameters are explained in detail in \citet{arras2022}.

\begin{table}[h]
\caption{Prior parameters for the Gaussian process models for the SCM and the MCM components} 
\label{tab:params}
\centering
\begin{tabular}{l c c c c}
\hline\hline
Model & SCM && MCM \\
& mean & std & mean & std \\
\hline
    zero mode &  1 & 1 &  1 & 1 \\
    fluctuations & 5 & 1&  5 & 1  \\
    loglogavgslope & -2 & 0.5&  -2 & 0.5 \\
    flexibility & 1.2 & 0.4 &  1.2 & 0.4 \\
    asperity & 0.2 & 0.2 &  0.2 & 0.2 \\
\hline
\end{tabular}
\end{table}

\section{Additional plots}

This section contains additional figures that were excluded from
the main text to save space, but that we nevertheless want to
include for completeness.

\begin{figure}[h]
  \resizebox{\hsize}{!}{\includegraphics{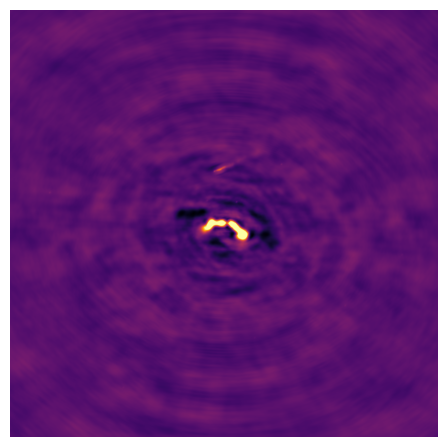}}
  \caption{Natural weighted dirty image of the ESO137-006 MeerKAT observation in the LO sub-band (961–1145 MHz) in units of Jy/beam.}
  \label{fig:eso_dirty}
\end{figure}

\Cref{fig:eso_dirty} depicts the natural weighted dirty image of the ESO137-006 MeerKAT observation. It serves as a starting point for the inference with \texttt{aim-resolve}. 

\begin{figure*}
\sidecaption
  \includegraphics[width=12cm]{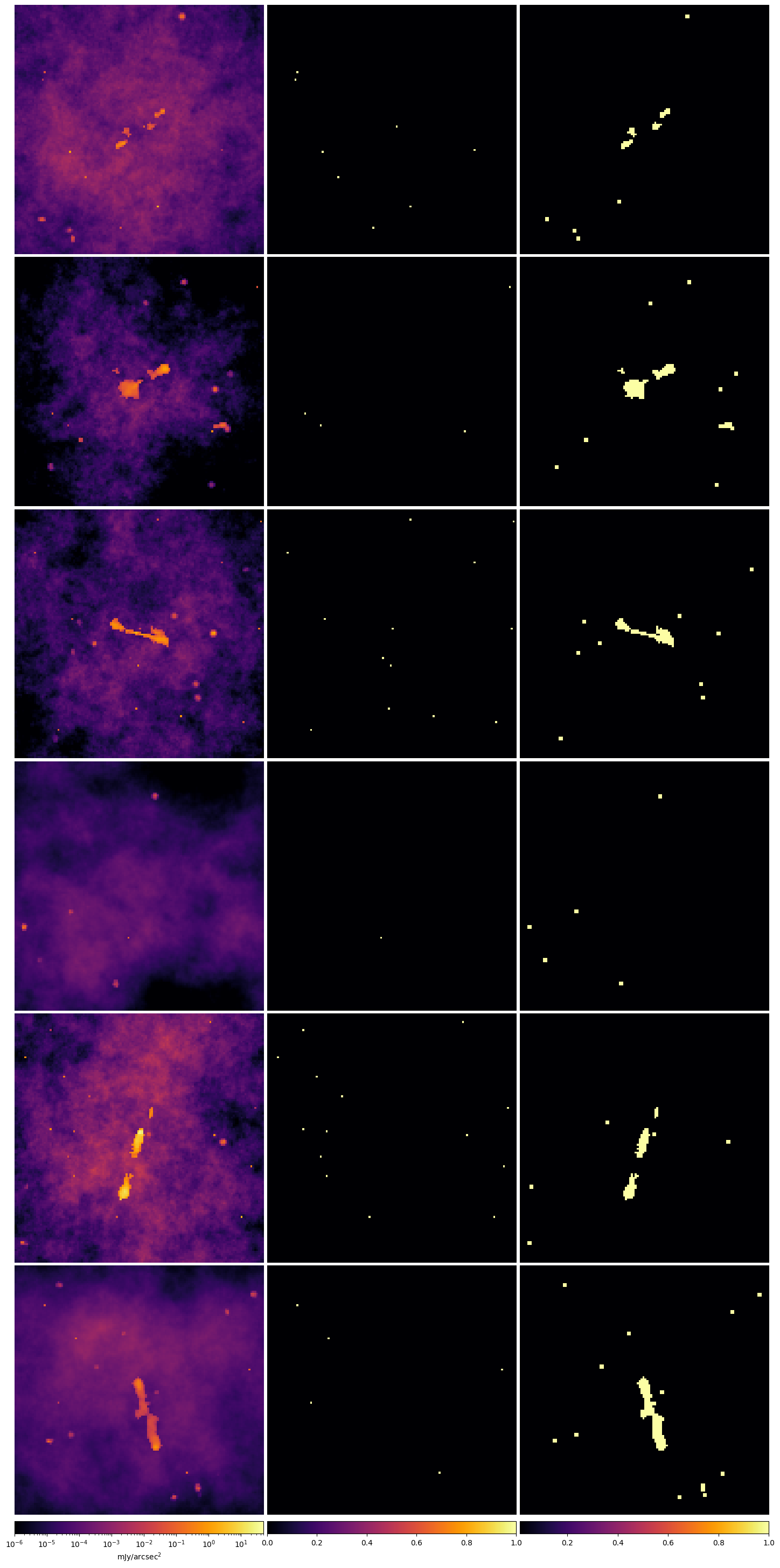}
     \caption{Subset containing six samples of the train-128 data used to train the U-Net. Each sample consists of an input image and two segmentation masks, one for point sources and one for extended objects (from left to right).}
     \label{fig:train_data}
\end{figure*}

\Cref{fig:train_data} shows a subset of the train-128 data used to train the U-Net. The corresponding validation and test datasets contain similar images and labels but use distinct sets of galaxy-like objects from the \texttt{RadioGalaxyDataset}. The same holds for the datasets with resolutions of $512 \times 512$ and $1024 \times 1024$.

\begin{figure*}
\sidecaption
  \includegraphics[width=12cm]{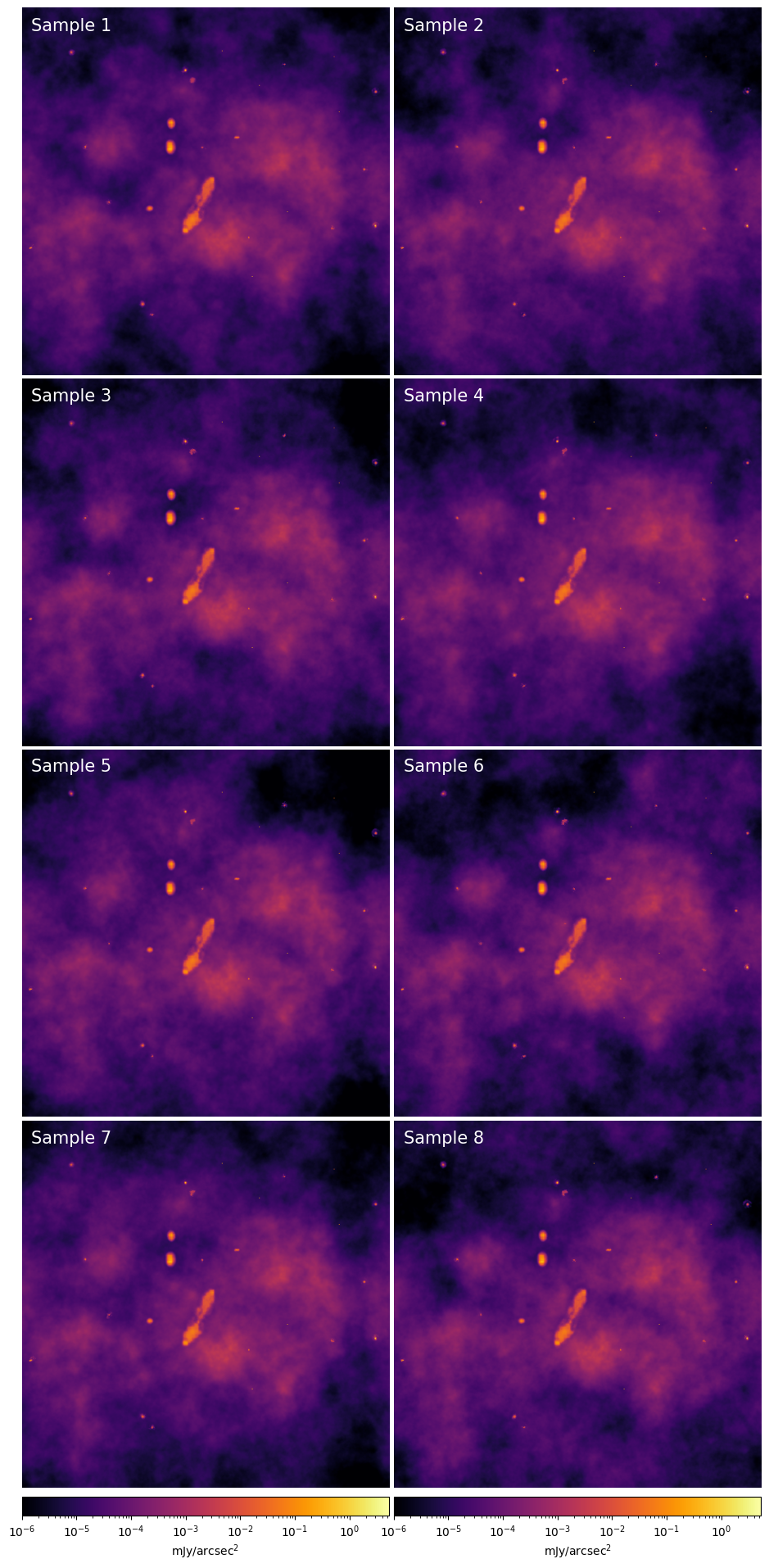}
     \caption{Reconstructed posterior samples of the synthetic radio data using the MCM (after the 3rd iteration of aim-resolve).}
     \label{fig:exp_samples}
\end{figure*}

\begin{figure*}
\sidecaption
  \includegraphics[width=12cm]{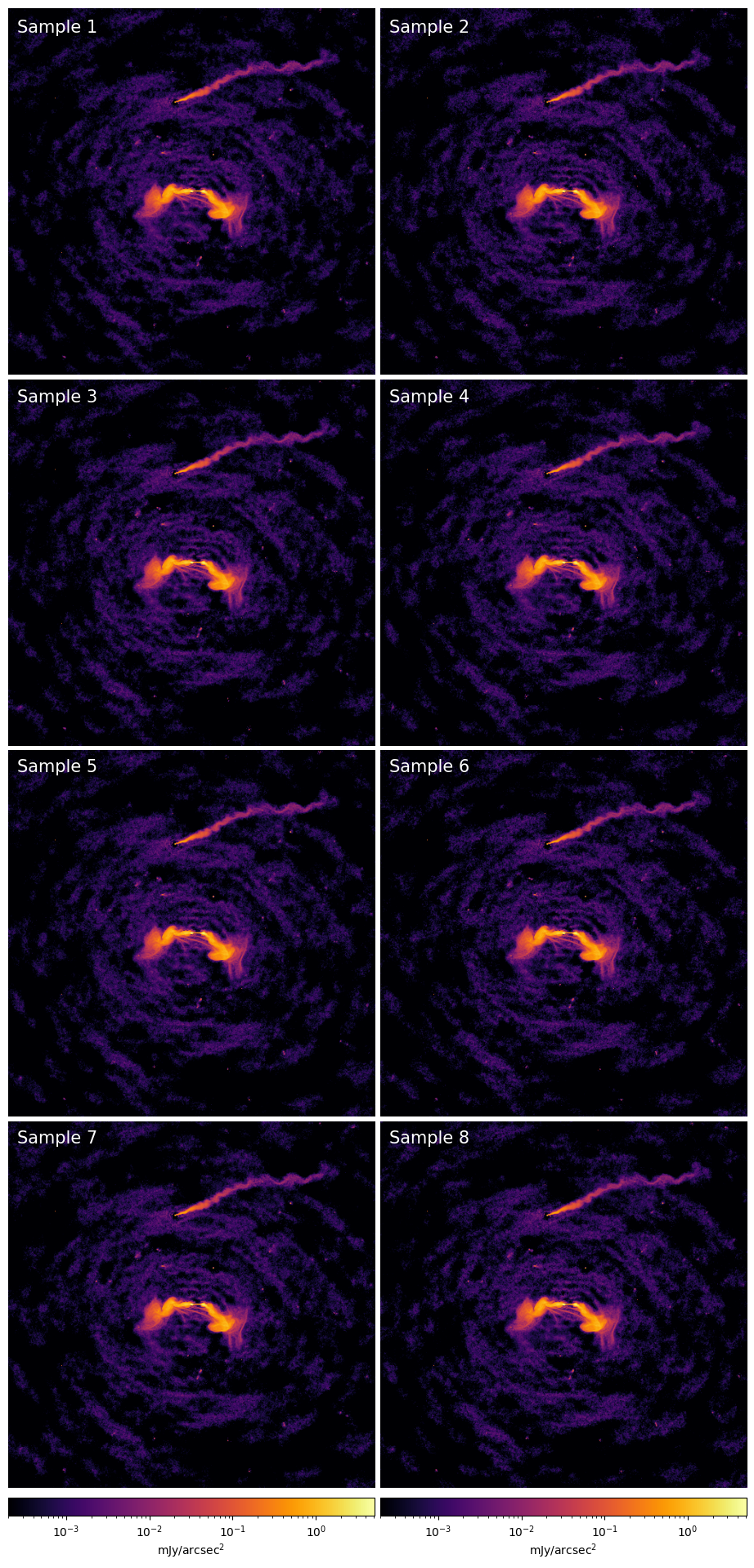}
     \caption{Reconstructed posterior samples (central $1 \degr \times 1 \degr$) of the ESO137-006 MeerKAT observation in the LO sub-band (961–1145 MHz) using the MCM (after the 2nd iteration of aim-resolve).}
     \label{fig:eso_samples}
\end{figure*}

\Cref{fig:exp_samples} and \cref{fig:eso_samples} illustrate the reconstructed MCM posterior samples of the synthetic radio data and the ESO137-006 MeerKAT observation, respectively. These samples are used to build summary statistics of the approximate posterior distribution like the reconstructed posterior mean or the relative standard deviation that are shown in the applications section. The individual samples (e.g. in \cref{fig:exp_samples}) show differences especially in parts of the diffuse background, indicating high uncertainty in these areas. The opposite holds for the point sources and extended objects, resulting in low uncertainties at their locations.
 
\label{LastPage}

\end{document}